\documentclass[letterpaper, 10 pt, journal, twoside]{IEEEtran}  % Comment this line out if you need a4paper
\pdfoutput=1

\IEEEoverridecommandlockouts                              % This command is only needed if 
\usepackage{graphics} % for pdf, bitmapped graphics files
\usepackage{epsfig} % for postscript graphics files
\usepackage{times} % assumes new font selection scheme installed
\usepackage{amsmath} % assumes amsmath package installed
\usepackage{amssymb}  % assumes amsmath package installed
\usepackage{hyperref}
\usepackage{multirow}
\usepackage{algorithm}
\usepackage{amssymb}
\usepackage{amsmath,amsfonts}
\usepackage{amsopn}
\usepackage{bm} % bold symbol
\usepackage{multirow}

\newcommand{\vct}[1]{\boldsymbol{#1}} % vector
\newcommand{\mat}[1]{\boldsymbol{#1}} % matrix
\newcommand{\field}[1]{\mathbb{#1}}
\newcommand{\R}{\field{R}} % real domain
\newcommand{\N}{\field{N}}
\newcommand{\C}{\field{C}} % complex domain
\newcommand{\ProbOpr}[1]{\mathbb{#1}}
\newcommand{\expect}[2]{%
\ifthenelse{\equal{#2}{}}{\ProbOpr{E}_{#1}}
{\ifthenelse{\equal{#1}{}}{\ProbOpr{E}\left[#2\right]}{\ProbOpr{E}_{#1}\left[#2\right]}}} % Expectation: syntax: E{1}{2} = E_1[2], E{}{2}=E[2], E{1}{} = E_1
\newcommand{\var}[2]{%
\ifthenelse{\equal{#2}{}}{\ProbOpr{VAR}_{#1}}
{\ifthenelse{\equal{#1}{}}{\ProbOpr{VAR}\left[#2\right]}{\ProbOpr{VAR}_{#1}\left[#2\right]}}} % Expectation: syntax: V{1}{2} = V_1[2], V{}{2}=V[2], V{1}{} = V_1
\newcommand{\vp}{\vct{p}}

\newcommand{\vx}{{\vct{x}}}

\newcommand{\vz}{{\vct{z}}}

\newcommand{\mG}{\mat{G}}

\newcommand{\eat}[1]{}

\def\etal{\MakeLowercase{\textit{et al.}}}
\def\lidar{lidar}
\def\Lidar{Lidar}
\def\namelong{Label Diffusion \Lidar{} Segmentation}
\def\nameshort{LDLS}

\title{
LDLS: 3D Object Segmentation through Label Diffusion from 2D Images 
}

\author{Brian H. Wang$^{1}$, Wei-Lun Chao$^{2}$, Yan Wang$^{2}$, Bharath Hariharan$^{2}$, Kilian Q. Weinberger$^{2}$, and Mark Campbell$^{1}$%
\thanks{Manuscript received: February 24th, 2019; Revised May 16th, 2019; Accepted May 24th, 2019.}%Use only for final RAL version
\thanks{This paper was recommended for publication by Editor Cesar Cadena upon evaluation of the Associate Editor and Reviewers' comments. This work was supported by grants from the Office of Naval Research (N00014-17-1-2175), and from the National Science Foundation (Smart \& Autonomous Systems Program IIS-1724282, and TRIPODS 1740822). We are grateful for the support of SAP and Facebook Research.}%Use only for final RAL version
\thanks{$^{1}$Brian H. Wang and Mark Campbell are with the Sibley School of Mechanical and Aerospace Engineering, Cornell University, Ithaca, NY 14850, USA.
{\tt\small \{bhw45, mc288\}@cornell.edu}}%
\thanks{$^{2} $Wei-Lun Chao, Yan Wang, Bharath Hariharan, and Kilian Q. Weinberger are with the Department of Computer Science, Cornell University, Ithaca, NY 14850, USA.
{\tt\small \{wc635, yw763, bh497, kqw4\}@cornell.edu}}%
\thanks{Digital Object Identifier (DOI): 10.1109/LRA.2019.2922582}
}
\begin{document}

\maketitle

%%%%%%%%%%%%%%%%%%%%%%%%%%%%%%%%%%%%%%%%%%%%%%%%%%%%%%%%%%%%%%%%%%%%%%%%%%%%%%%%

%%%%%%%%%%%%%%%%%%%%%%%%%%%%%%%%%%%%%%%%%%%%%%%%%%%%%%%%%%%%%%%%%%%%%%%%%%%%%%%%

\begin{abstract}
Object segmentation in 3D point clouds is a critical task for robots capable of 3D perception. Despite the impressive performance of deep learning-based approaches on object segmentation in 2D images, deep learning has not been applied nearly as successfully for 3D point cloud segmentation. Deep networks generally require large amounts of labeled training data, which are readily available for 2D images but are difficult to produce for 3D point clouds. In this paper, we present \namelong{} (\nameshort{}), a novel approach for 3D point cloud segmentation which leverages 2D segmentation of an RGB image from an aligned camera to avoid the need for training on annotated 3D data. We obtain 2D segmentation predictions by applying Mask-RCNN to the RGB image, and then link this image to a 3D \lidar{} point cloud by building a graph of connections among 3D points and 2D pixels. This graph then directs a semi-supervised label diffusion process, where the 2D pixels act as source nodes that diffuse object label information through the 3D point cloud, resulting in a complete 3D point cloud segmentation. We conduct empirical studies on the KITTI benchmark data set and on a mobile robot, demonstrating wide applicability and superior performance of \nameshort{} compared to the previous state-of-the-art in 3D point cloud segmentation, without any need for either 3D training data or fine-tuning of the 2D image segmentation model.
\end{abstract}
% Keywords appear just beneath the abstract. Use only for final RAL version.
\begin{IEEEkeywords}
Object Detection, Segmentation and Categorization; RGB-D Perception
\end{IEEEkeywords}
\section{INTRODUCTION}
\label{S_intro}

\begin{figure}
	\includegraphics[width=\columnwidth]{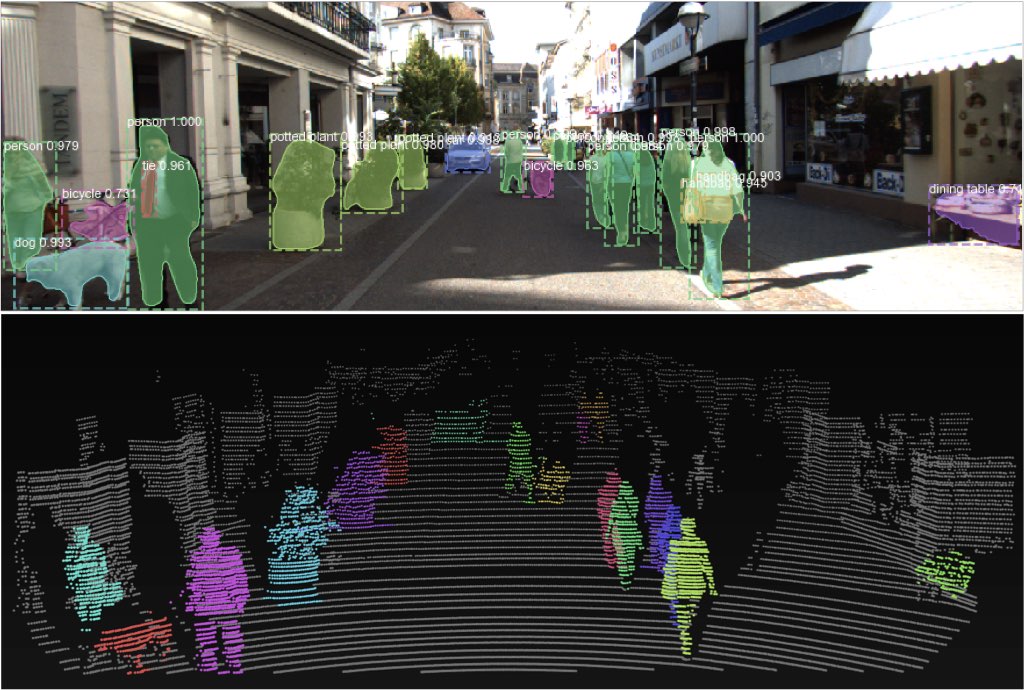}
	\vskip -5pt
	\caption{Our approach, \namelong~(\nameshort), performs instance segmentation on 3D \lidar{} point clouds by leveraging information from an aligned camera. This allows \nameshort{} to recognize a wide variety of object classes (including people, bicycles, a car, and a dog, in this example image), while requiring no training on labeled 3D \lidar{} data.}
	\label{fig:segmentation_example}
	\vskip -10pt
\end{figure}

% Drop letter for first word of the Introduction
\IEEEPARstart{R}{obots} in a variety of applications require the ability to recognize objects of interest in their environment and distinguish them from the background, using 3D sensors. Examples range from autonomous cars detecting nearby pedestrians, to an industrial robot identifying an object to be assembled. 
Modern \lidar{} sensors and stereo cameras allow robots to perceive their 3D surroundings in detail as a point cloud. In particular, \lidar{} sensors provide 3D spatial measurements with extreme precision. 
For a robot to effectively navigate and/or interact with its surroundings, 
a method is needed to reliably perform detection and point-by-point segmentation of  objects in 3D point clouds. Figure \ref{fig:segmentation_example} demonstrates this task. 

Recently, object detection and segmentation in 2D images have undergone impressive improvements in accuracy and reliability through the rise of approaches based on deep convolutional neural networks~\cite{He2017,Liu2016,Ren2015}. 2D images are a natural fit for deep learning: their inherent pixel grid structure allows convolutional neural networks to be applied effectively and efficiently~\cite{Krizhevsky2012}, and large data sets can be labeled easily by non-experts, enabling deep neural networks to thrive. 

In contrast, 3D point clouds are unstructured by nature and laborious to label. Although some successful attempts have been made to convert point clouds into inputs for deep networks~\cite{Qi2017,wu2018squeezeseg,wu2018squeezesegv2,wang2018pointseg,Zhou2017,Ku2017,Qi2018}, the lack of large labeled data sets serves as an inherent limitation. Annotating 3D point clouds is a time-consuming and difficult process; a single \lidar{} point cloud can contain tens of thousands of points, and labeling of many such point clouds would be required. Furthermore, these challenges scale with the number of object classes, and with the variety in environments. As evidence for the asymmetry between the difficulty of 2D and 3D labeling, one can look at the relative sizes of the MS COCO~\cite{Lin2014} and KITTI~\cite{Geiger2012} data sets - the image-only MS COCO data set contains over 200,000 labeled images and 1.5 million labeled object instances, whereas the image-and-\lidar{} KITTI object detection data set contains 7,481 image-point cloud pairs for training and 7,518 for testing, with 80,256 object instances.

In this paper we propose a novel approach to 3D object segmentation, which bypasses the challenges inherent in the use of 3D point clouds with deep neural networks. Our method leverages the success of convolutional neural networks for 2D image segmentation and frames 3D point cloud segmentation as a semi-supervised learning problem on graphs~\cite{Zhu2002,zhu2005semi}. 
In order to leverage the advantages of both data modalities, we observe a scene with a 3D \lidar{} sensor \emph{and} an aligned 2D camera. We apply an off-the-shelf object segmentation algorithm (Mask-RCNN~\cite{He2017}) to the 2D image in order to detect object classes and instances at the pixel-by-pixel level.
Subsequently, we construct a graph by connecting 2D pixels to 3D \lidar{} points according to their 2D projected locations, as well as connecting \lidar{} points that neighbor one another in 3D space. We use label diffusion~\cite{Zhu2002} to propagate 2D segmentation labels through this graph, thereby labeling the 3D \lidar{} points.
We refer to our algorithm as \emph{\namelong{} (\nameshort{}}).
The end result is a fully labeled 3D point cloud, obtained by leveraging the strengths of both data modalities. The camera provides rich semantic information about object classes and facilitates the application of deep neural networks, and the \lidar{} sensor provides precise 3D spatial measurements. Additionally, by identifying objects in 3D at the point level, rather than outputting rectangular 3D bounding boxes around detected objects~\cite{Zhou2017,Yang2018,Ku2017,Qi2018,Du2018}, \nameshort{} allows much more precise localization of object instances which do not neatly fit into rectangular boxes, such as people, animals, etc. 

Although labeled images are required to train the 2D object segmentation model, this is a much lower barrier than requiring 3D annotated point clouds. Furthermore, open-source pretrained models are available for a variety of object classes; thus collection of training data may not even be necessary in some cases. By removing the need for labeled 3D training data, \nameshort{} facilitates generalization to new environments, object classes, and robot sensor configurations.

We conduct experiments on the KITTI data set~\cite{Geiger2012,Geiger2013IJRR} to quantitatively evaluate and analyze the accuracy of \nameshort{}. Our experimental results include evaluations on a subset of the KITTI object detection data set which we have manually annotated with point-level labels, in order to evaluate point-wise segmentation. We additionally present qualitative evaluations on KITTI data, as well as on data collected with a mobile robot equipped with a camera and lower-resolution 3D \lidar{} sensor. Our results demonstrate that \nameshort{} can be applied successfully in different domains on different sensors, and is capable of detecting a variety of object classes due to the flexibility of the pretrained image object detector. Our manually labeled KITTI ground truth data set, and an open-source implementation of \nameshort{}, are shared publicly at \url{https://github.com/brian-h-wang/LDLS}.

\section{RELATED WORK}
\label{S_related}

\subsection{Deep Learning on Point Clouds}

Deep neural networks have been proposed for various perception tasks on point clouds in the past. PointNet~\cite{Qi2017} defines a network architecture that operates directly on unstructured point clouds and extracts features that are invariant to point re-ordering, capturing both local and global point cloud information. PointNet and its successor PointNet++~\cite{Qi2017a} have been shown to be successful at point cloud classification and semantic segmentation, and have also been extended to object detection~\cite{Qi2018} and instance segmentation~\cite{wang2018sgpn}.

Other methods extend convolutional neural networks to point clouds. Since 3D points lack the grid structure of images, one approach is to arrange the points into a 3D voxel grid and perform 3D convolution~\cite{Zhou2017}; however this can be computationally inefficient, especially for sparse \lidar{} point clouds~\cite{Ren2018}. Alternatively, points can be projected into 2D, using panoramic projection~\cite{wu2018squeezeseg, wang2018pointseg} or a bird's-eye view~\cite{Yang2018}. These projections allow 2D convolution, but information is lost in the reduction to 2D, rendering these approaches unsuitable for some environments, especially complex or cluttered scenes. Recently, \cite{wang2018deep} and \cite{Li2018} propose direct convolution over point clouds by adjusting the kernel weights locally according to irregular point positions.

Several methods for object detection in autonomous driving scenes use \lidar{} sensor data alongside images from an aligned camera~\cite{Ku2017,Qi2018,Du2018}, using the KITTI data set for training and evaluation~\cite{Geiger2012}. 
While the KITTI data set is an excellent benchmark, its annotations are 3D bounding boxes, and object classes are limited to driving-relevant objects such as pedestrians, cars, and cyclists. There is significant potential benefit to a variety of robotics applications in building methods that recognize a broader variety of objects, and output point-level segmentations. Along these lines, a key limitation---and opportunity---in deep learning on point clouds is the expansion of labeled data. Recent works~\cite{Lee2018,lertniphonphan20182d} attempt to ease the process of annotating ground truth point clouds. 

\subsection{Graphical Models and 2D-3D Fusion}

As an alternative to deep neural networks, graphical models have been successfully applied to point clouds for various tasks. Previous works by Maddern \& Newman~\cite{Maddern2016} and Schoenberg \etal~\cite{Schoenberg2010} use graphical models to fuse \lidar{} scans and stereo camera depth maps to produce accurate dense depth maps suitable for use on autonomous vehicles. Wang \etal~\cite{Wang2013}~propose a semantic segmentation method for image-aligned 3D point clouds by retrieving referenced labeled images of similar appearances and then propagating their labels to the 3D points using a graphical model.

Various other approaches for fusion of 2D and 3D information have also been considered. Wang \& Neumann~\cite{Wang2018DepthCNN} add depth-aware operations to standard CNNs to improve segmentation performance for RGBD images. Xie \etal{}~\cite{Xie2016} developed a method to rapidly annotate 2D street scenes by first drawing labeled bounding primitives in 3D, and then transferring labels on to 2D images. Zhang \etal{}~\cite{Zhang2018} train a neural network for 2D semantic segmentation, then project onto dense 3D data from a long-range laser scanner. This work uses the additional assumption of coplanar dense points sharing labels to clean projected labels and achieve semantic background segmentation of classes such as buildings and roads. These various concepts of 2D-3D fusion demonstrate that point clouds and images can be used to complement one another in the perception process.

\subsection{Learning with Graphs}

Our proposed approach is based on semi-supervised learning with graphs. Graph-based methods are well-established in machine learning, especially for nearest-neighbor graphs built upon the manifold assumption~\cite{belkin2004regularization, zhu2005semi}. In this paper, we construct the graph according to 3D \lidar{} point locations, as well as their projected 2D image pixel coordinates, and adapt the label diffusion algorithm by Zhu~\cite{zhu2005semi}.

\section{APPROACH}

% diffusion graph matrix
\def\G{\mG}
% Submatrices for G:
%	Upper-left: Lidar-to-lidar / 3D-to-3D
%	Upper-right: Image-to-lidar / 2D-to-3D
\def\Gl{\G^{3D\rightarrow3D}}
\def\Gr{\G^{2D\rightarrow3D}}

\def\pixeltopointweight{\lambda}

% Projection set notation
\def\proj{\mathcal{P}}

% The number of 3D lidar points
\def\npt{N_{points}}
% The number of 2D image pixels
\def\npx{N_{pixels}}
% The total number of nodes (n_points + n_pixels)
\def\ntot{N}

\def\labelvec{z^{(m)}}

\label{S_approach}

\begin{figure}
	\centering
	\includegraphics[width=\columnwidth]{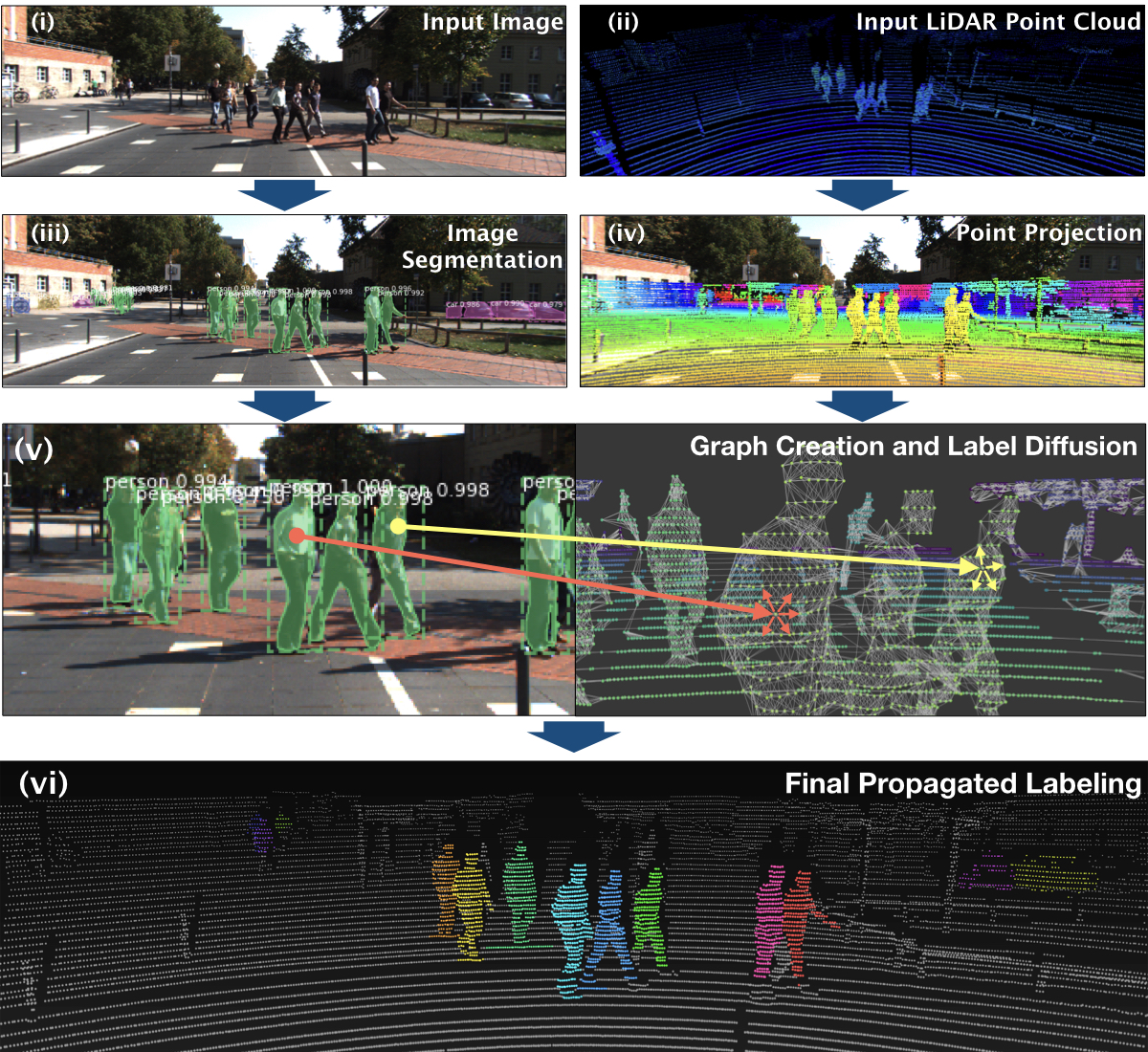}
	\vskip -5pt
	\caption{The full segmentation pipeline, from the input point cloud and image to the final \lidar{} point cloud segmentation.}\label{fig:pipeline}
	\vskip -10pt
\end{figure}

This section presents our label diffusion method for object instance segmentation in \lidar{} point clouds. We formulate the task as a semi-supervised learning problem on a graph, and leverage 2D segmentation from an RGB image along with 3D geometry from a point cloud to obtain a complete 3D segmentation. Figure \ref{fig:pipeline} shows the segmentation pipeline.

\subsection{Problem Formulation}

Object instance segmentation in 3D point clouds can be formulated as follows: Given a point cloud $\{\vx_i\in\R^3\}_{i=1}^{\npt}$, where $\npt$ is the number of 3D points, we want to assign to every point $\vx_i$ an object instance label $y_i\in \{0\}\cup\N$. Each instance is also associated with a class label $c\in\C$, where $\C=\{0,\cdots,C\}$ is the label set. Let $y=0$, $c=0$ indicate the background instance and class.

Our approach takes in as inputs a \lidar{} point cloud $\{\vx_i\in\R^3\}_{i=1}^{\npt}$ and an aligned RGB image. This is a common sensor configuration for autonomous vehicles and mobile robots. Note that we consider only \lidar{} points which lie within the field of view of the camera. The pixels have 2D coordinates defined as $\{\vp_i\in\N^2\}_{i=1}^{\npx}$, where $\npx$ is the total number of pixels in the image.  Algorithms for 2D object instance segmentation on RGB images have been well-developed (e.g., Mask R-CNN~\cite{He2017}) and pretrained models on large-scale data sets (e.g., MS COCO~\cite{Lin2014}) are readily accessible for a wide variety of object classes. We therefore leverage these resources for the task of labeling \lidar{} points.

We approach the task of 3D segmentation from the perspective of semi-supervised learning, to avoid training a segmentation model on annotated point clouds. Semi-supervised learning assumes that a set of data points is available, of which a subset of points is labeled. For graph-based semi-supervised learning, we label the remaining points by defining connections between data points and then diffusing labels along these connections~\cite{zhu2005semi}. To apply this framework to \lidar{} point cloud segmentation, we construct a graph by drawing connections from 2D pixels to 3D \lidar{} points, as well as among the 3D points. The 2D pixels are labeled according to results from 2D object segmentation of the RGB image, and the graph is then used to diffuse labels onto the 3D points, which are all initially unlabeled. The success of this approach depends on appropriate choices for the graph structure, and label diffusion process.

\subsection{Graph Construction}
The graph $\G$ used in our method consists of two types of nodes (2D image pixels, and 3D \lidar{} points), as well as two types of connections between nodes (from a 2D pixel to a 3D point, and between two 3D points). 

\paragraph{Initial Graph Node Labeling}  Each 2D pixel and each 3D point is a node within $\G$. Initially, all 3D points are unlabeled. The 2D pixels are labeled according to an image segmentation algorithm, which assigns every image pixel an instance label $y$ (where $y=0$ corresponds to the background), and associates each instance with a class label $c$. The output will therefore be several distinct instance masks, each containing many pixels, as seen in Figure \ref{fig:pipeline}~(\emph{iii}). This instance-class association is deterministic for each image, simplifying the task of assigning instance labels to the \lidar{} points within the camera's field of view.

\paragraph{2D-to-3D Connections} Since we assume the camera and \lidar{} sensors are aligned, each \lidar{} point can be projected from 3D into 2D image pixel coordinates, as shown in Figure \ref{fig:pipeline}~(\emph{iv}). At this stage, a naive way to label the \lidar{} point cloud would be to label each point according to the 2D instance mask into which it is projected. This method will however result in significant labeling errors, especially around the 2D instance boundaries, due to calibration errors between the sensors as well as the fact that the 2D segmentation masks are unaware of 3D depth. Background \lidar{} points will therefore often project into a foreground object mask, or vice versa.  Our segmentation pipeline should therefore combine 2D and 3D information, and leverage both information sources for producing a final 3D segmentation.

In order to combine 2D and 3D information in our graph for semi-supervised label diffusion, we construct a subgraph $\Gr$ connecting 2D pixels to 3D \lidar{} points, represented by a $(\npt \times \npx)$ matrix. 
\begin{align}
\Gr_{ij} = 
\begin{cases}
\pixeltopointweight & \text{if } \vp_j \in \proj(\vx_i)\\
0, & \text{otherwise.}
\end{cases}\label{graph_image2lidar}
\end{align}
$\proj(\vx_i)$ is the set of image pixels which are near the projected 2D location of \lidar{} point $\vx_i$. In our implementation, $\proj(\vx_i)$ is the set of all pixels in a 5 pixel by 5 pixel box centered around the projected 2D coordinates of $\vx_i$. The parameter $\pixeltopointweight$ controls the amount of information that can flow from a pixel to a connected \lidar{} point. In our experiments, we use a small constant value of $\pixeltopointweight=0.001$, to mitigate sensor calibration errors by minimizing the influence of any one pixel. While more sophisticated schemes like setting different box sizes and $\pixeltopointweight$ values for each $\vx_i$ may be applied, empirically we find our design choice to perform well across multiple domains.

\paragraph{3D-to-3D Connections}
In order to encode connections between 3D points, we construct a nearest neighbor graph from the points to reflect the underlying 3D geometry. This subgraph is denoted as $\Gl$, represented by a $(\npt \times \npt)$ matrix.

Given a \lidar{} point cloud $\{\vx_i\in\R^3\}_{i=1}^{\npt}$, we construct an exponential-weighted nearest neighbors graph over the points. For each point $\vx_i$, we compute $KNN(\vx_i)$, the set of $K$ nearest neighbors to $\vx_i$ within the point cloud, according to Euclidean distance. The graph of 3D point connections is then defined as
\begin{align}
\Gl_{ij} = 
\begin{cases}
1, & \text{if } i=j\text{, else}\\
\exp \left(-\frac{\| \vx_i - \vx_j \|^2_2}{\sigma}\right), & \text{if } \vx_j \in KNN(\vx_i)\\
0, & \text{otherwise.}
\end{cases}\label{graph_lidar2lidar}
\end{align}
Each nonzero element $\Gl_{ij}$ captures the similarity between points $\vx_i$ and $\vx_j$. With a small $K$, this subgraph is sparse, enabling fast computation during the diffusion step later on---we set $K=10$ and $\sigma=1$ in our experiments. We apply a KD-tree to speed up the construction of $\Gl$.

\paragraph{Full Label Diffusion Graph}

The full graph for label diffusion, combining the 2D-to-3D connections as well as the 3D-to-3D connections, is then defined as
\begin{equation}\label{eq:graph}
	\G = \begin{bmatrix}
		\Gl & \Gr \\
		\mat{0} & \mat{I}
	\end{bmatrix},
\end{equation}
where $\mat{I}$ is the $(\npx\times\npx)$ identity matrix. Let $\ntot=\npt+\npx$; then $\G$ is $(N \times N)$. This graph is illustrated in Figure \ref{fig:pipeline}~(\emph{v}).
After constructing $\G$ according to Eq. (\ref{eq:graph}), we normalize each row of the matrix to sum to 1, following \cite{Zhu2002}.
\begin{align}
\G_{ij} \leftarrow \cfrac{\G_{ij}}{\sum_{j'}\G_{ij'}.} \label{e_normalize}
\end{align}

\vskip -15pt
\begin{algorithm}[b]
	\caption{\nameshort: \namelong}	\label{algo}
	\textbf{Required:} A 2D instance segmentation algorithm, and a 3D-to-2D projection matrix.\\
	\textbf{Input:} A 3D point cloud $\{\vx_i\}_{i=1}^{\npt}$ and an aligned image.\\
	\textbf{Graph Creation:} \\
	1: Perform 2D segmentation to generate instance masks.\\
	2. Project 3D points into 2D and connect them with pixels.\\
	3. Connect each 3D point with its $K$ nearest neighbor points.\\
	4. Construct the graph matrix $\G$ (cf. Eq.~(\ref{graph_image2lidar})-(\ref{e_normalize})).\\
	\textbf{Label Diffusion:} \\
	5. Define a label vector $\vz^{(m)}$ for each instance (cf. Eq.~(\ref{e_label})). Set initial labels for 3D points to zero, and set labels for 2D pixels based on 2D segmentation masks.\\
	6. Perform label diffusion (cf. Eq.~(\ref{e_LD})) until convergence, or the maximum number of iterations is achieved.\\
	7. Determine instance labels $\{y_i\}_{i=1}^{\npt}$ (cf. Eq.~(\ref{e_output})).\\
	8. Remove outliers (cf. Eq~(\ref{outliers})) and return final point labels.
\end{algorithm}

\subsection{Label Diffusion} 
The graph matrix $\G$ guides the label diffusion process for \lidar{} point labeling. The nonzero elements of $\G$ indicate connections along which information on object instance labels should be diffused. The intuition behind the diffusion process is for the 2D pixels to act as source nodes that continuously push label information out through the 3D points, which is then diffused throughout the point cloud according to the connections between points. This process cleans up segmentation boundaries using 3D geometry information, resulting in a final 3D segmentation incorporating both 2D and 3D information.

To perform label diffusion, let us assume in total $M+1$ object instances $\{0,1,\cdots,M\}$, including the background instance, are detected by the 2D segmentation method (the background instance can be defined by the absence of any object mask). Let us define $\vz^{(m)}\in\R^N$ as an $N$-dimensional label vector for instance $m$. $\vz^{(m)}$ contains one entry for each 3D point and 2D pixel. The entries corresponding to 3D points are initialized to zero, and the entries corresponding to the 2D pixels are defined according to the 2D segmentation masks:
\begin{align}
\vz^{(m)}_{i}=
\begin{cases}
0, & \text{if } i < \npt\text{, else}\\
\mu(m, i-\npt), & \text{otherwise,}
% 
% \npt < i \leq \npx \text{and } \vx_i \text{ is seeded with instance $m$},\\
% 0, & \text{otherwise.} 
\end{cases}
\label{e_label}
\end{align}
where the mask function $\mu(m,j)$ returns 1 if pixel $\vp_j$ is in the segmentation mask of object instance $m$, and 0 otherwise. 

We then iteratively perform the following computation,
\begin{align}
\vz^{(m)} \leftarrow \G \times \vz^{(m)} \label{e_LD}
\end{align}
to diffuse labels throughout the graph nodes, for all $M+1$ instances. Note that if point $\vx_i$ is unlabeled, but connected to at least one pixel $\vp_j$ labeled with instance $m$ and $\Gr_{ij}>0$, then after such a computation we obtain $z^{(m)}_i>0$, indicating an increased likelihood that $\vx_i$ will be labeled with instance $m$ as a result of label diffusion from $\vp_j$.

Note that the construction of $G$ with $\begin{bmatrix}
	\mat{0} & \mat{I}
\end{bmatrix}$ as the bottom submatrix ensures that the pixel labels in $\vz^{(m)}$ remain unchanged by this matrix multiplication. Since the labels of all initially labeled nodes within the graph remain fixed, and since $\G$ is row-normalized, the label diffusion is proven to converge according to Zhu~\cite{zhu2005semi}.

We iteratively apply label diffusion according to Eq.~(\ref{e_LD}) until convergence of all $\vz^{(m)}$, or until a maximum number of iterations (200, in our experiments). Finally, we then convert the likelihood values to \lidar{} point labels according to
\begin{equation}
y_i = \text{argmax}_{m\in\{0,1,\cdots,M\}} z^{(m)}_i. \label{e_output}
\end{equation}
That is, we assign each point the most likely label.

Label diffusion can sometimes result in disjointed sections of object segmentations; most often this occurs if projection or mask boundary errors result in a large number of contiguous background \lidar{} points being projected to inside a 2D segmentation mask. To clean up these errors, we introduce an outlier removal step based on finding connected components within $\G$. Let $\G^{(m)}$ be the subgraph of $\Gl$ defined by considering only \lidar{} points labeled as object $m$, i.e. $\vx_i$ is a node of $\G^{(m)}$ if and only if $y_i=m$. Then, let $\mathcal{C}(\G^{m})$ be the largest connected component in $\G^{(m)}$, treating it as an undirected graph. We update the \lidar{} point labels as
\begin{align}
y_{i} \leftarrow
\begin{cases}
y_i, & \text{if } \vx_i \in \mathcal{C}(\G^{m})\\
0, & \text{otherwise.}
% 
% \npt < i \leq \npx \text{and } \vx_i \text{ is seeded with instance $m$},\\
% 0, & \text{otherwise.} 
\end{cases}
\forall y_i \in \{y|y=m\}
\label{outliers}
\end{align}
The final output of this pipeline is a \lidar{} point cloud where each point is labeled as either a background point, or as part of an object instance (with a class label). Algorithm~\ref{algo} summarizes our overall algorithm.

\section{RESULTS AND EVALUATION}
\label{S_exp}

\subsection{Quantitative Evaluation on the KITTI Data Set}
We present experiments on the KITTI data set~\cite{Geiger2012} in order to quantitatively study the accuracy of \nameshort{}. Our evaluation considers the \emph{Pedestrian} and \emph{Car} object classes, as these are \emph{a)} the most common object classes in the KITTI data set and \emph{b)} appear in both the KITTI and MS COCO data sets. This allows us to generate results using Mask-RCNN~\cite{He2017} pretrained on MS COCO, and then evaluate on KITTI. In most of our experiments, we merge the \emph{Pedestrian} and \emph{Person\_sitting} KITTI classes. The KITTI object detection leaderboard also includes cyclists, however since the Mask-RCNN model we use is trained to detect people and bicycles separately, we do not consider this class. We perform no training or fine-tuning of Mask-RCNN on KITTI data, allowing us to study the applicability of our method to a new domain unseen in the image training data.

\subsubsection{Performance Metrics}

\def\c{\mathcal{C}}
\def\pred{\mathcal{P}}
\def\gt{\mathcal{G}}
\def\predc{\pred_c}
\def\gtc{\gt_c}
\def\match{\mathcal{M}}

We evaluate using performance metrics for both semantic (per-class) segmentation and instance segmentation~\cite{wu2018squeezeseg}. Given a set of \lidar{} points, semantic segmentation is defined as assigning a class label $c \in \C$ to each point, where $\C$ is the set of all classes. In our experiments, $\C = \{Car, Pedestrian\}$. Given point class labels, the segmentation is evaluated by computing precision, recall, and IoU over all $n$ points for each class $c$, as
\begin{equation}\label{eq:sem_precision}
	P_{c} = \frac{|\predc \cap \gtc|}{|\predc|},
	R_{c} = \frac{|\predc \cap \gtc|}{|\gtc|},
	IoU_{c} = \frac{\vert \predc \cap \gtc\vert}{\vert \predc \cup \gtc \vert},
\end{equation}
where $\predc$ is the set of \lidar{} points predicted to have class $c$ in the segmentation results, $\gtc$ is the set of points with class $c$ in the ground truth, and $|\cdot|$ denotes cardinality of a set.

An instance segmentation additionally assigns each point an instance label, distinguishing individual cars and pedestrians from one another. To evaluate an instance segmentation, predicted instance labels $i\in\{1,\dots,M_P\}$ must be matched with corresponding ground truth instances $j\in\{1,\dots,M_G\}$. We do this by calculating the IoU between each prediction-ground truth instance pair with the same class label, and then calculating a bipartite graph matching $\match(i)=j$ which maximizes the sum of IoUs between matched instance pairs.

The number of true positives $TP$ is then calculated by counting the number of prediction-truth matchings with IoU over some predefined threshold. The false positive and false negative counts $FP$ and $FN$ are determined by the number of unmatched prediction instances and truth instances, respectively, and instance precision and recall are then computed as
\begin{equation}
	P = \frac{TP}{TP+FP}, \text{ }R = \frac{TP}{TP+FN}.
\end{equation}
It is also possible to calculate instance precision and recall by counting over individual points, rather than instances. However for \lidar{} data, this metric is unevenly weighted towards objects that are closer to the sensor because they possess a higher density of points. Our results report instance-level precision and recall in order to avoid this bias.

\subsubsection{Comparison to Other \Lidar{} Segmentation Methods}
\def\numannotated{200}

\begin{table*}
\caption{Comparison of semantic segmentation accuracy between SqueezeSeg~\cite{wu2018squeezeseg}, SqueezeSegV2~\cite{wu2018squeezesegv2}, PointSeg~\cite{wang2018pointseg}, and \nameshort{}.}\label{table:comparison}
\begin{center}
\vskip-10pt
\begin{tabular}{c c |c c c | c c c}
& &  \multicolumn{3}{c}{\textbf{Noisy Ground Truth (2791 Frames)}} & \multicolumn{3}{c}{\textbf{Manually Labeled Ground Truth (200 Frames)}} \\
Class & Method & Precision & Recall & IoU & Precision & Recall & IoU\\
\hline
\multirow{4}{*}{Car} & SqueezeSeg & 66.7 & 95.4 & 64.6 & 51.0 & \textbf{97.2} & 50.3\\
 & SqueezeSegV2 & 81.7 & 87.5 & 73.2 & 63.7 & 90.2 & 59.6\\
 & PointSeg & 77.2 & \textbf{96.2} & \textbf{74.9} &  58.6 & 89.8 & 54.9\\
 & \emph{\nameshort{}} & \textbf{87.4} & 75.0 & 67.7 & \textbf{84.1} & 88.4 & \textbf{75.7} \\
 \hline
& SqueezeSeg & 52.9 & 28.6 & 22.8 & 49.7 & 29.7 & 22.8 \\
Pedestrian & SqueezeSegV2 & \textbf{57.4} & 35.0 & 27.8 & 71.1 & 36.2 & 31.5 \\
 \textit{(without Person\_sitting)}& PointSeg & 48.6 & 29.4 & 22.4 & 70.6 & 20.4 & 18.8\\
 & \emph{\nameshort{}} & 51.3 & \textbf{81.9} & \textbf{46.0} & \textbf{75.6} & \textbf{87.3} & \textbf{68.1} \\
\hline
\end{tabular}
\end{center}
\vskip-10pt
\end{table*}

We benchmark \nameshort{} through a comparison against SqueezeSeg~\cite{wu2018squeezeseg}, SqueezeSegV2~\cite{wu2018squeezesegv2}, and PointSeg~\cite{wang2018pointseg}, state-of-the-art convolutional neural network methods for object segmentation in \lidar{} point clouds. 
These methods take as input a \lidar{} point cloud transformed through panoramic projection into a $64\times512\times5$ tensor, where the 5 channels are $x$-, $y$-, and $z$-coordinates, depth, and \lidar{} intensity. Since KITTI provides 3D bounding box object annotations, rather than point labels, Wu \etal~\cite{wu2018squeezeseg} generated segmentation annotations by labeling points that fall within the annotated 3D boxes, producing an 8057-frame training set and a 2791-frame validation set.

However, this bounding box-based annotation process results in labeling errors, for example from extra points that fall within the boxes. In order to maximize the quality of our evaluation, we therefore manually annotated \numannotated{} KITTI point clouds (randomly selected from within the validation set used by SqueezeSeg) with class and instance ground truth segmentations. In our annotation process, point labels are initialized according to the KITTI bounding boxes, and then manually cleaned up by annotators. The end result is a set of pristinely labeled ground truth point clouds. Note that KITTI objects whose bounding boxes contain no \lidar{} points are dropped. In order to quantify the error of the bounding box-generated point labels, we calculated the semantic segmentation IoU of the KITTI box-derived labels with our manually annotated labels, and found an overall IoU of 81.1 for cars and 93.4 for pedestrians, indicating a significant amount of error in bounding box-generated labels.

We apply \nameshort{} to the 2791-frame SqueezeSeg validation data set, as well as our \numannotated{}-frame manually labeled data set. These two evaluations each present a different takeaway. The 2791-frame evaluation is useful due to its large scale, and due to its usage by previous works. However, due to the bias caused by annotation errors, this evaluation cannot definitively establish real-world segmentation performance. Rather, its purpose is to broadly establish competitiveness of our method alongside existing works. The smaller-scale evaluation on manual annotations complements the first evaluation by removing annotation error, and therefore offering the clearest possible indicator of expected real-world performance.

Table \ref{table:comparison} presents results from these two evaluations. SqueezeSeg, SqueezeSegV2, and PointSeg results are produced after training on the 8057-point cloud training set. For these methods, results on the noisy ground truth are taken from their respective papers~\cite{wu2018squeezeseg,wu2018squeezesegv2,wang2018pointseg}. As the SqueezeSeg validation data set does not include instance labels, Table \ref{table:comparison} presents results using semantic segmentation metrics only. For consistency with the other methods, we treat \emph{Pedestrian} and \emph{Person\_sitting} as two separate classes in this experiment only, even though the Mask-RCNN model used in \nameshort{} does not distinguish between the two. This only slightly impacts our results, due to the small number of \emph{Person\_sitting} instances in KITTI. Results in all other following experiments are presented with these two classes merged.

When measured on the SqueezeSeg validation data, \nameshort{} achieves competitive semantic segmentation performance compared to SqueezeSeg, SqueezeSegV2, and PointSeg, although without any training on labeled 3D data. Additionally, \nameshort{} outperforms the other methods in overall pedestrian segmentation IoU. On the manually labeled data, the difference is far more pronounced. Our method achieves a $27.0$\% increase in IoU for car segmentation and a $116.2$\% increase for pedestrian segmentation as compared to the next-best method. 
We hypothesize that one reason for the difference in performance between \nameshort{} and SqueezeSeg/PointSeg is that the latter methods were trained on data that includes annotation errors. When the test set includes similar annotation errors and is therefore closer to the training domain (i.e. the SqueezeSeg validation set), the performance difference between \nameshort{} and the other methods is smaller. However, this gap grows wider for evaluation on error-free annotations.

It is noted that this comparison is not symmetrical---SqueezeSeg and PointSeg make use of labeled 3D training data, while \nameshort{} uses RGB images as well as a pretrained 2D segmentation model (implicitly also using 2D image training data). However, it is important to view this difference in the context of robotics applications. \Lidar{}-equipped robots commonly also have a camera sensor, or can be inexpensively outfitted with one. On the other hand, creating a labeled training set of thousands of \lidar{} point clouds is difficult and time-consuming. Therefore, if installing a camera on a robot and then applying an off-the-shelf image segmentation model can enable similar or greater 3D segmentation performance as a model trained on labeled point clouds, adding the camera information is a reasonable choice in a robotics context.

\subsubsection{Instance Segmentation Evaluation}

We also present an instance segmentation evaluation by applying \nameshort{} to our manually annotated KITTI ground truth data. Results calculated across all object instances in the ground truth are shown in Table \ref{table:instance_seg}.
\begin{table}
\caption{Instance segmentation performance on manual annotations.}\label{table:instance_seg}
\begin{center}
\vskip-10pt
\begin{tabular}{c c|c c c c c}
IoU Threshold & Class & Precision & Recall & TP & FP & FN \\
 \hline
\multirow{2}{*}{0.50} & Car & 66.8 & 79.3 & 544 & 270 & 142  \\
 & Pedestrian & 51.4 & 68.4 & 128 & 121 & 59  \\
\hline
\multirow{2}{*}{0.70} & Car & 57.7 & 68.5 & 470 & 344 & 216 \\
 & Pedestrian & 48.6 & 64.7 & 121 & 128 & 66 \\
\hline
\end{tabular}
\end{center}
\vskip-10pt
\end{table}
To better understand the performance of \nameshort{}, we additionally study the effect of object range on accuracy. Label diffusion assumes that neighboring points are more likely to share a class label; this assumption weakens at further distances from the sensor, where \lidar{} data becomes sparser. Therefore, we hypothesize that \nameshort{} should be more reliable for objects that are closer to the sensor and visible with a higher density of points.
\begin{figure}
  \centering
  \includegraphics[width=\columnwidth]{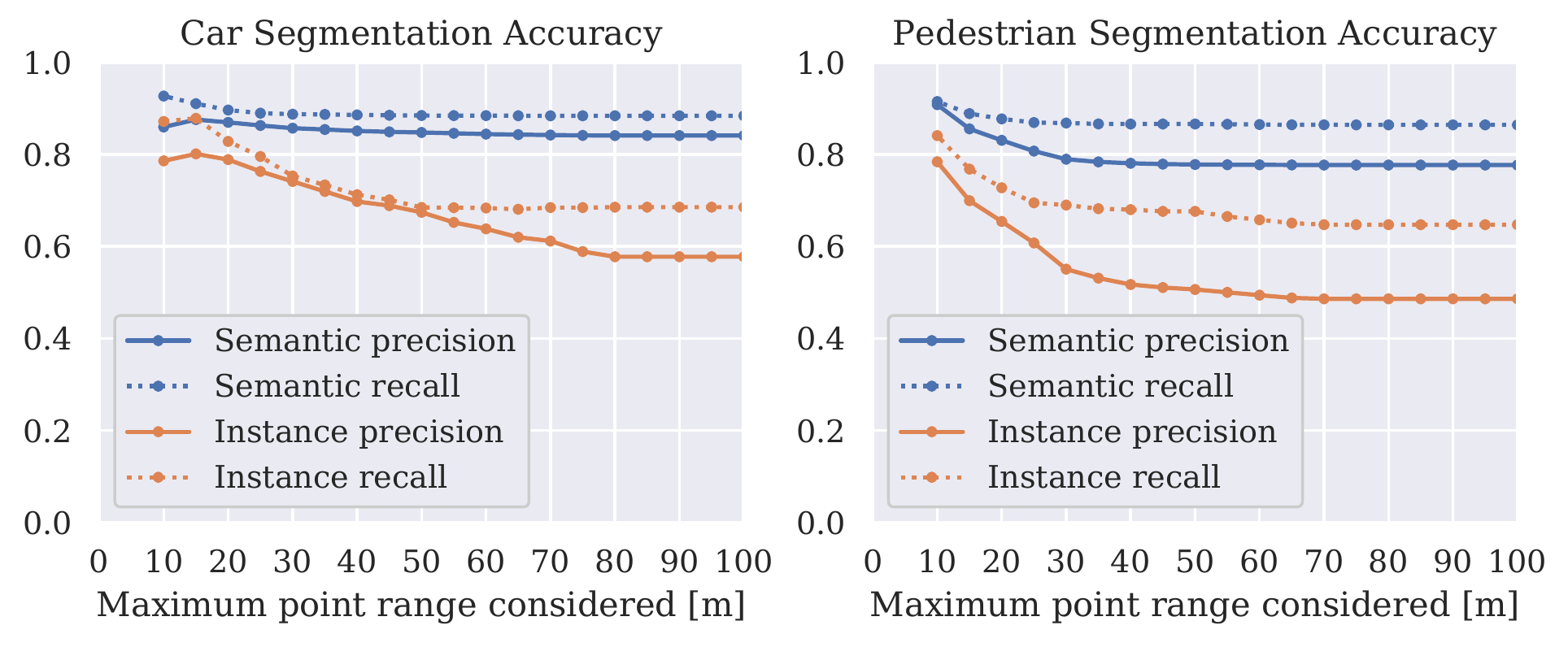}
  \vskip-10pt
  \caption{Effect of range on semantic and instance segmentation precision and recall. Instance segmentation metrics use $IoU=.70$.}\label{fig:range_vs_pr}
  \vskip-10pt
\end{figure}
\begin{figure}[t]
	\centering
  \includegraphics[width=\columnwidth]{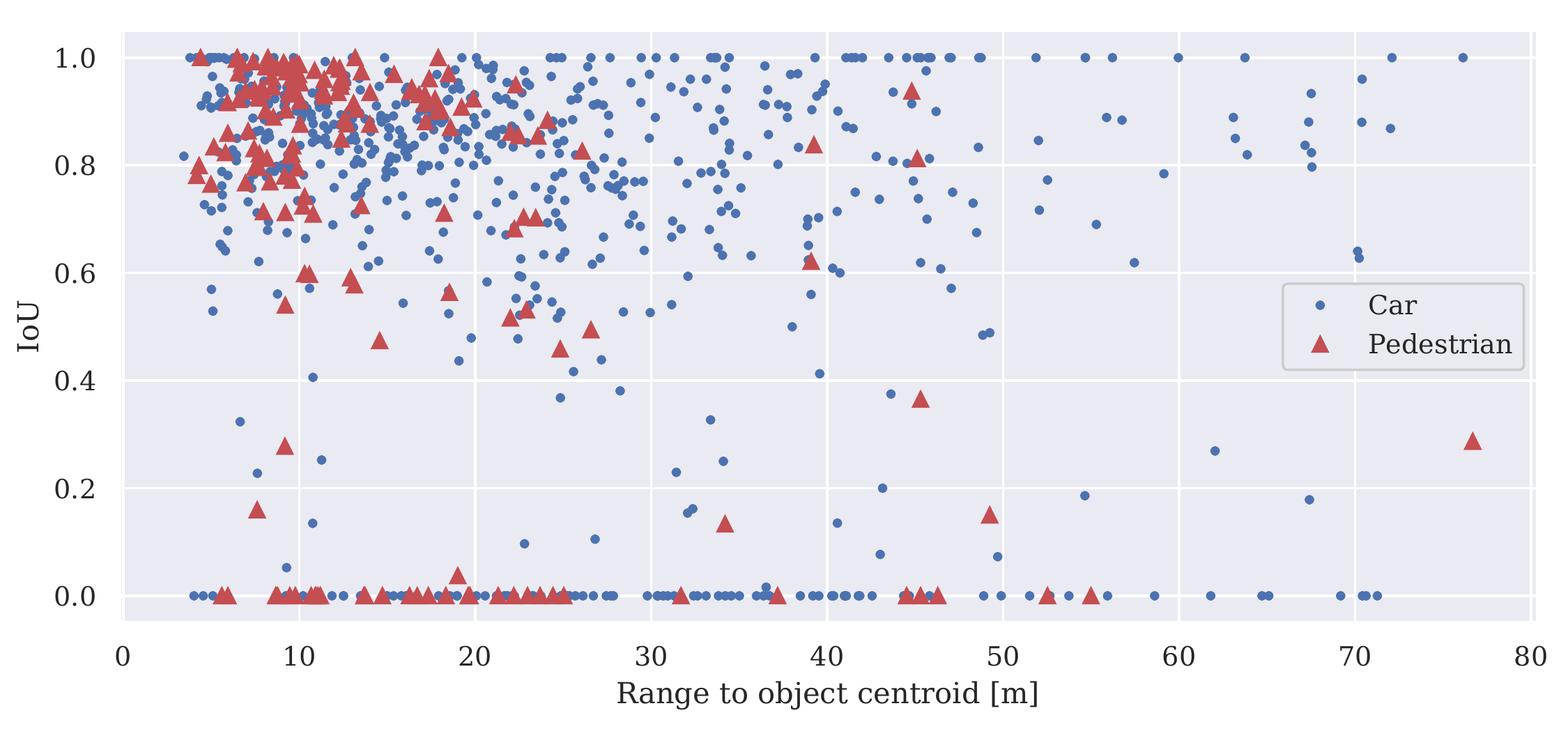}
  \vskip-10pt
  \caption{Scatter plot showing object range versus segmentation IoU. Each point is a pedestrian or car instance. Zero IoU points indicate false negatives.}\label{fig:range_vs_iou_scatter}
  \vskip-10pt
\end{figure}
We test this hypothesis by performing evaluations at different ranges. For each evaluation, we exclude all \lidar{} points above a given maximum range cutoff. Results are plotted in Figure \ref{fig:range_vs_pr}. As range from the sensor increases, instance segmentation performance decreases significantly. Semantic segmentation degradation is not as significant; likely because these metrics count numbers of points and therefore are biased towards nearer, more densely populated \lidar{} points.

Figure \ref{fig:range_vs_iou_scatter} plots each object instance within our test set on a scatter plot, as a function of distance to the object's centroid against instance segmentation IoU. As objects become more distant, a wider range of IoU results appear.

These experiments indicate that \nameshort{} generally segments object instances more reliably at closer distances, with performance falling off as range increases. This suggests that an all-purpose robotic perception system may be best served by using a vision-based bounding box object detector at far ranges, with \nameshort{} applied at close ranges to allow a robot to precisely sense and interact with its immediate surroundings.

\subsubsection{Ablation Study}

To demonstrate the benefits of the different components of the \nameshort{}  pipeline, we perform an ablation study by removing different components and comparing results. The ablation settings we experimented with are: 

\begin{enumerate}
  \item \emph{Direct projection labeling} \Lidar{} points are naively labeled, without graph diffusion, based on whether they project to within a 2D segmentation mask in the image. 
  \item \emph{Diffusion without outlier removal} The full pipeline is executed, except for the final outlier removal step. 
\end{enumerate}

Table \ref{table:ablation} contains semantic and instance (using both the 0.50 and 0.70 IoU thresholds) segmentation results from running these settings on the manually annotated data. We see that both the diffusion and outlier removal steps improve overall performance, with the former contributing a major performance gain. This finding confirms the value of label diffusion in fusing 2D and 3D information. Note that the full pipeline results differ slightly from the results presented in Table \ref{table:comparison}, since the ablation results are calculated after merging the \emph{Person\_sitting} and \emph{Pedestrian} classes.

\begin{table*}
\begin{center}
\caption{Ablation study results.}\label{table:ablation}
\vskip -8pt
\tabcolsep 3.3pt
\begin{tabular}{c c|c c c|c c | c c}
& & \multicolumn{3}{c}{Semantic Segmentation} & \multicolumn{2}{c}{Instance Seg. IoU=0.50} & \multicolumn{2}{c}{Instance Seg. IoU=0.70}\\
Ablation Setting & Class & Precision & Recall & IoU & Precision & Recall & Precision & Recall \\
\hline
\multirow{2}{*}{Direct projection labeling without diffusion} & Car & 69.2 & 83.2 & 60.7 & 44.1 & 59.8 & 16.4 & 22.2\\
& Pedestrian + Person Sitting & 60.1 & 72.2 & 48.8 & 30.0 & 51.9 & 5.3 & 9.1\\
\hline
\multirow{2}{*}{Diffusion without outlier removal} & Car & 78.2 & 90.8 & 72.5 & 66.3 & 78.7 & 51.6 & 61.2\\
& Pedestrian & 72.9 & 89.1 & 66.9 & 50.6 & 67.4 & 47.4 & 63.1\\
 \hline
\multirow{2}{*}{Original, with all components} & Car & 84.1 & 88.4 & 75.7 & 66.8 & 79.3 & 57.7 & 68.5\\
& Pedestrian & 77.7 & 86.4 & 69.2 & 51.4 & 68.4 & 48.6 & 64.7\\
\hline
\end{tabular}
\end{center}
\vskip-10pt
\end{table*}

\subsection{Qualitative Evaluation}
\begin{figure*}
	\centering
   \includegraphics[width=\textwidth]{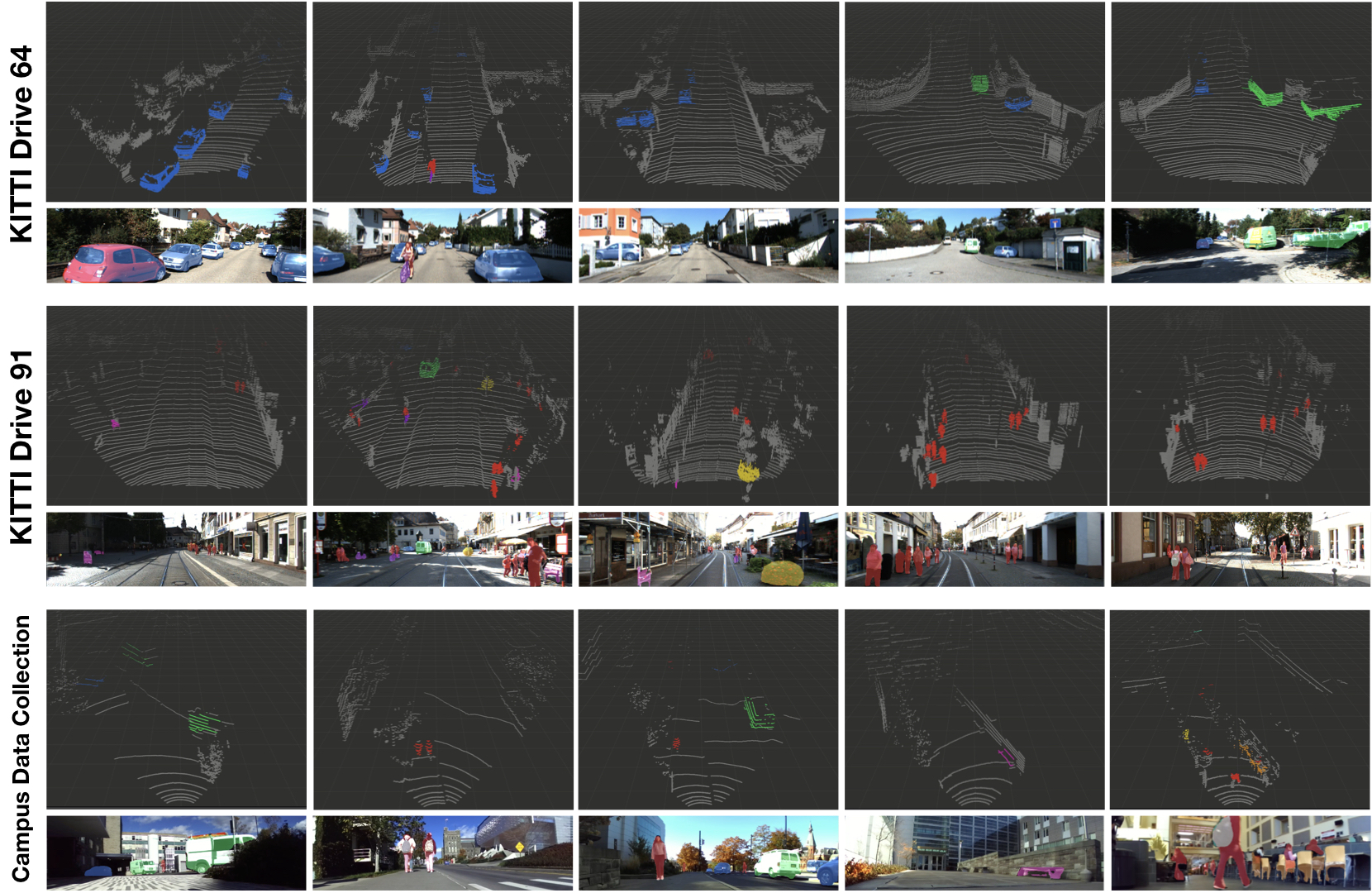}
   \vskip-10pt
   \caption{Qualitative results from running \nameshort{} on the KITTI Drive 64 and Drive 91 sequences, as well as on data collected on the Cornell campus using a mobile robot. Highlighted COCO object classes segmented in the \lidar{} data are 'person' (shown in red), 'car' (blue), 'truck' (green), 'bicycle' (purple), 'chair' (orange), 'bench' (pink), and 'potted plant' (yellow). Mask-RCNN segmentations are shown in the corresponding images.}\label{fig:qual_eval}
   \vskip-10pt
\end{figure*}
Our quantitative study establishes the accuracy of \nameshort{} on large-scale annotated ground truth, but is limited to just two object classes, and data from a high-resolution \lidar{} sensor. In order to complement the quantitative evaluation, and demonstrate the applicability of \nameshort{} to different environments, classes, sensors, and data collection platforms, we present two additional sets of results: \emph{a)} residential and urban sequences from KITTI~\cite{Geiger2013IJRR}, and \emph{b)} a sequence captured on the Cornell University campus using a Clearpath\texttrademark{} Jackal mobile ground robot with a Velodyne VLP-16 \lidar{} sensor and RGB camera. 

For this evaluation, we present segmentation results on a wider variety of MS COCO classes, including people, cars, trucks, bicycles, chairs, benches, and potted plants. Segmentation of all of these object classes in \lidar{} data is made possible since \nameshort{} adopts all classes that are detected by the image segmentation model used; in this case the Mask-RCNN model pretrained on the MS COCO data set. An overview of results is shown in Figure \ref{fig:qual_eval}. In the KITTI data, new object classes are generally segmented with qualitatively comparable accuracy to cars and pedestrians, although narrower objects such as bicycles present a challenge. In comparison, the campus data collection on the Jackal robot exhibits more segmentation errors, and performance breaks down more significantly for objects at farther distances. We hypothesize the following reasons for these differences: Firstly, the VLP-16 sensor outputs only 16 laser scan lines, as opposed to the 64-scan \lidar{} used in KITTI, making the \lidar{} point clouds sparser and more difficult to segment, especially at further ranges. We also believe that errors from sensor calibration~\cite{Geiger2012a} and time synchronization were higher on the Jackal, compared to the KITTI data set.  Still, we find \nameshort{} to demonstrate adequate segmentation performance for a small, relatively inexpensive robot.

In terms of computation, the graph construction and sparse matrix multiplication iterations are highly parallelizable and can be GPU-accelerated~\cite{Bell2008}. On the KITTI evaluation data, our current Python implementation averages approximately .38 seconds per frame on an Nvidia GTX 1080 Ti, excluding the computation of Mask-RCNN results.

%-------------------------------------------------------------------------

\section{CONCLUSION}

In this paper, we present \nameshort{}, a method for instance segmentation of 3D point clouds which leverages a pretrained 2D image segmentation model, followed by label diffusion on a graph connecting 2D pixels and 3D points, to remove any need for labeled 3D training data. By removing this requirement, we make \nameshort{} suitable for application in various environments and on different robotic platforms. Quantitative evaluations on the KITTI data set demonstrate superior accuracy at car and pedestrian segmentation compared to previous methods, and qualitative evaluations demonstrate segmentation of a wider variety of object classes. Our results are additionally presented without fine-tuning of the image segmentation model, demonstrating generalizability to new domains.

%%%%%%%%%%%%%%%%%%%%%%%%%%%%%%%%%%%%%%%%%%%%%%%%%%%%%%%%%%%%%%%%%%%%%%%%%%%%%%%%

%%%%%%%%%%%%%%%%%%%%%%%%%%%%%%%%%%%%%%%%%%%%%%%%%%%%%%%%%%%%%%%%%%%%%%%%%%%%%%%%

%%%%%%%%%%%%%%%%%%%%%%%%%%%%%%%%%%%%%%%%%%%%%%%%%%%%%%%%%%%%%%%%%%%%%%%%%%%%%%%%

\section*{ACKNOWLEDGMENT}
The authors would like to thank Sarah Allen, Christopher Graef, Emily Sun, and Shuo Han for their help with collecting and preparing the data used in the presented experiments.

%%%%%%%%%%%%%%%%%%%%%%%%%%%%%%%%%%%%%%%%%%%%%%%%%%%%%%%%%%%%%%%%%%%%%%%%%%%%%%%%

% References are important to the reader; therefore, each citation must be complete and correct. If at all possible, references should be commonly available publications.

% \bibliographystyle{IEEEtran}
% \bibliography{ReferencesIROS2019}

\begin{thebibliography}{10}
\providecommand{\url}[1]{#1}
\csname url@samestyle\endcsname
\providecommand{\newblock}{\relax}
\providecommand{\bibinfo}[2]{#2}
\providecommand{\BIBentrySTDinterwordspacing}{\spaceskip=0pt\relax}
\providecommand{\BIBentryALTinterwordstretchfactor}{4}
\providecommand{\BIBentryALTinterwordspacing}{\spaceskip=\fontdimen2\font plus
\BIBentryALTinterwordstretchfactor\fontdimen3\font minus
  \fontdimen4\font\relax}
\providecommand{\BIBforeignlanguage}[2]{{%
\expandafter\ifx\csname l@#1\endcsname\relax
\typeout{** WARNING: IEEEtran.bst: No hyphenation pattern has been}%
\typeout{** loaded for the language `#1'. Using the pattern for}%
\typeout{** the default language instead.}%
\else
\language=\csname l@#1\endcsname
\fi
#2}}
\providecommand{\BIBdecl}{\relax}
\BIBdecl

\bibitem{He2017}
K.~He, G.~Gkioxari, P.~Doll{\'a}r, and R.~Girshick, ``Mask r-cnn,'' in
  \emph{ICCV}, 2017.

\bibitem{Liu2016}
W.~Liu, D.~Anguelov, D.~Erhan, C.~Szegedy, S.~Reed, C.-Y. Fu, and A.~C. Berg,
  ``Ssd: Single shot multibox detector,'' in \emph{ECCV}, 2016.

\bibitem{Ren2015}
S.~Ren, K.~He, R.~Girshick, and J.~Sun, ``Faster r-cnn: Towards real-time
  object detection with region proposal networks,'' in \emph{NeurIPS}, 2015.

\bibitem{Krizhevsky2012}
A.~Krizhevsky, I.~Sutskever, and G.~E. Hinton, ``Imagenet classification with
  deep convolutional neural networks,'' in \emph{NeurIPS}, 2012.

\bibitem{Qi2017}
C.~R. Qi, H.~Su, K.~Mo, and L.~J. Guibas, ``Pointnet: Deep learning on point
  sets for 3d classification and segmentation,'' in \emph{CVPR}, 2017.

\bibitem{wu2018squeezeseg}
B.~Wu, A.~Wan, X.~Yue, and K.~Keutzer, ``Squeezeseg: Convolutional neural nets
  with recurrent crf for real-time road-object segmentation from 3d lidar point
  cloud,'' in \emph{ICRA}, 2018.

\bibitem{wu2018squeezesegv2}
B.~Wu, X.~Zhou, S.~Zhao, X.~Yue, and K.~Keutzer, ``Squeezesegv2: Improved model
  structure and unsupervised domain adaptation for road-object segmentation
  from a lidar point cloud,'' in \emph{ICRA}, 2019.

\bibitem{wang2018pointseg}
Y.~Wang, T.~Shi, P.~Yun, L.~Tai, and M.~Liu, ``Pointseg: Real-time semantic
  segmentation based on 3d lidar point cloud,'' \emph{arXiv preprint
  arXiv:1807.06288}, 2018.

\bibitem{Zhou2017}
Y.~Zhou and O.~Tuzel, ``Voxelnet: End-to-end learning for point cloud based 3d
  object detection,'' in \emph{CVPR}, 2018.

\bibitem{Ku2017}
J.~Ku, M.~Mozifian, J.~Lee, A.~Harakeh, and S.~L. Waslander, ``Joint 3d
  proposal generation and object detection from view aggregation,'' in
  \emph{IROS}, 2018.

\bibitem{Qi2018}
C.~R. Qi, W.~Liu, C.~Wu, H.~Su, and L.~J. Guibas, ``Frustum pointnets for 3d
  object detection from rgb-d data,'' in \emph{CVPR}, 2018.

\bibitem{Lin2014}
T.-Y. Lin, M.~Maire, S.~Belongie, J.~Hays, P.~Perona, D.~Ramanan,
  P.~Doll{\'a}r, and C.~L. Zitnick, ``Microsoft coco: Common objects in
  context,'' in \emph{ECCV}, 2014.

\bibitem{Geiger2012}
A.~Geiger, P.~Lenz, and R.~Urtasun, ``Are we ready for autonomous driving? the
  kitti vision benchmark suite,'' in \emph{CVPR}, 2012.

\bibitem{Zhu2002}
X.~Zhu and Z.~Ghahramani, ``Learning from labeled and unlabeled data with label
  propagation,'' Tech. Rep., 2002.

\bibitem{zhu2005semi}
X.~Zhu, J.~Lafferty, and R.~Rosenfeld, ``Semi-supervised learning with
  graphs,'' Ph.D. dissertation, Carnegie Mellon University, 2005.

\bibitem{Yang2018}
B.~Yang, W.~Luo, and R.~Urtasun, ``Pixor: Real-time 3d object detection from
  point clouds,'' in \emph{CVPR}, 2018.

\bibitem{Du2018}
X.~Du, M.~H. Ang, S.~Karaman, and D.~Rus, ``A general pipeline for 3d detection
  of vehicles,'' in \emph{ICRA}, 2018.

\bibitem{Geiger2013IJRR}
A.~Geiger, P.~Lenz, C.~Stiller, and R.~Urtasun, ``Vision meets robotics: The
  kitti dataset,'' \emph{International Journal of Robotics Research (IJRR)},
  2013.

\bibitem{Qi2017a}
C.~R. Qi, L.~Yi, H.~Su, and L.~J. Guibas, ``Pointnet++: Deep hierarchical
  feature learning on point sets in a metric space,'' in \emph{NeurIPS}, 2017.

\bibitem{wang2018sgpn}
W.~Wang, R.~Yu, Q.~Huang, and U.~Neumann, ``Sgpn: Similarity group proposal
  network for 3d point cloud instance segmentation,'' in \emph{CVPR}, 2018.

\bibitem{Ren2018}
M.~Ren, A.~Pokrovsky, B.~Yang, and R.~Urtasun, ``Sbnet: Sparse blocks network
  for fast inference,'' in \emph{CVPR}, 2018.

\bibitem{wang2018deep}
S.~Wang, S.~Suo, W.-C. Ma, A.~Pokrovsky, and R.~Urtasun, ``Deep parametric
  continuous convolutional neural networks,'' in \emph{CVPR}, 2018.

\bibitem{Li2018}
Y.~Li, R.~Bu, and X.~Di, ``{PointCNN: Convolution On X-Transformed Points},''
  in \emph{NeurIPS}, 2018.

\bibitem{Lee2018}
J.~Lee, S.~Walsh, A.~Harakeh, and S.~L. Waslander, ``Leveraging pre-trained 3d
  object detection models for fast ground truth generation,'' in \emph{ITSC},
  2018.

\bibitem{lertniphonphan20182d}
K.~Lertniphonphan, S.~Komorita, K.~Tasaka, and H.~Yanagihara, ``2d to 3d label
  propagation for object detection in point cloud,'' in \emph{ICME Workshops},
  2018.

\bibitem{Maddern2016}
W.~Maddern and P.~Newman, ``Real-time probabilistic fusion of sparse 3d lidar
  and dense stereo,'' in \emph{IROS}, 2016.

\bibitem{Schoenberg2010}
J.~R. Schoenberg, A.~Nathan, and M.~Campbell, ``Segmentation of dense range
  information in complex urban scenes,'' in \emph{IROS}, 2010.

\bibitem{Wang2013}
Y.~Wang, R.~Ji, and S.-F. Chang, ``Label propagation from imagenet to 3d point
  clouds,'' in \emph{CVPR}, 2013.

\bibitem{Wang2018DepthCNN}
W.~Wang and U.~Neumann, ``{Depth-Aware CNN for RGB-D Segmentation},'' in
  \emph{ECCV}, 2018.

\bibitem{Xie2016}
J.~Xie, M.~Kiefel, M.-T. Sun, and A.~Geiger, ``Semantic instance annotation of
  street scenes by 3d to 2d label transfer,'' in \emph{CVPR}, 2016.

\bibitem{Zhang2018}
R.~Zhang, G.~Li, M.~Li, and L.~Wang, ``{Fusion of images and point clouds for
  the semantic segmentation of large-scale 3D scenes based on deep learning},''
  \emph{ISPRS Journal of Photogrammetry and Remote Sensing}, 2018.

\bibitem{belkin2004regularization}
M.~Belkin, I.~Matveeva, and P.~Niyogi, ``Regularization and semi-supervised
  learning on large graphs,'' in \emph{COLT}, 2004.

\bibitem{Geiger2012a}
A.~Geiger, F.~Moosmann, {\"O}.~Car, and B.~Schuster, ``Automatic camera and
  range sensor calibration using a single shot,'' in \emph{ICRA}, 2012.

\bibitem{Bell2008}
N.~Bell and M.~Garland, ``{Efficient Sparse Matrix-Vector Multiplication on
  CUDA},'' Tech. Rep., 2008.

\end{thebibliography}


\end{document}